\documentclass[conference]{IEEEtran}
\IEEEoverridecommandlockouts

\usepackage{subfig,url}
\usepackage{cite}
\usepackage{amsmath,amssymb,amsfonts}
\usepackage{algorithmic}
\usepackage{graphicx}
\usepackage{textcomp}
\usepackage{xcolor,balance}
\usepackage{nicefrac}
\usepackage{flushend}
\usepackage{calc}
\def\BibTeX{{\rm B\kern-.05em{\sc i\kern-.025em b}\kern-.08em
    T\kern-.1667em\lower.7ex\hbox{E}\kern-.125emX}}
\begin{document}

\title{Benchmarking Dynamic SLO Compliance in Distributed Computing Continuum Systems}
\author{
\IEEEauthorblockN{Alfreds Lapkovskis\IEEEauthorrefmark{1}, Boris Sedlak\IEEEauthorrefmark{2}, Sindri Magnússon\IEEEauthorrefmark{1}, Schahram Dustdar\IEEEauthorrefmark{2}\IEEEauthorrefmark{3}, and Praveen Kumar Donta\IEEEauthorrefmark{1}}\\
\IEEEauthorblockA{\IEEEauthorrefmark{1}\textit{Department of Computer Systems and Sciences (DSV)}, \textit{Stockholm University,} SE-106 91 Stockholm, Sweden }
\texttt{\{alfreds.lapkovskis, praveen, sindri.magnusson\}@dsv.su.se}
\IEEEauthorblockA{\IEEEauthorrefmark{2}\textit{Distributed Systems Group, TU Wien, Vienna 1040, Austria} \\
 \texttt{\{b.sedlak,dustdar\}@dsg.tuwien.ac.at}}
\IEEEauthorblockA{\IEEEauthorrefmark{3}\textit{ICREA, Universitat Pompeu Fabra Barcelona,} Barcelona 08002, Spain}
}

\maketitle

\begin{abstract}
Ensuring Service Level Objectives (SLOs) in large-scale architectures, such as Distributed Computing Continuum Systems (DCCS), is challenging due to their heterogeneous nature and varying service requirements across different devices and applications. 
Additionally, unpredictable workloads and resource limitations lead to fluctuating performance and violated SLOs. 
To improve SLO compliance in DCCS, one possibility is to apply machine learning; however, the design choices are often left to the developer. To that extent, we provide a benchmark of Active Inference—an emerging method from neuroscience—against three established reinforcement learning algorithms (Deep Q-Network, Advantage Actor-Critic, and Proximal Policy Optimization).
We consider a realistic DCCS use case: an edge device running a video conferencing application alongside a WebSocket server streaming videos. 
Using one of the respective algorithms, we continuously monitor key performance metrics, such as latency and bandwidth usage, to dynamically adjust parameters—including the number of streams, frame rate, and resolution—to optimize service quality and user experience. 
To test algorithms' adaptability to constant system changes, we simulate dynamically changing SLOs and both instant and gradual data-shift scenarios, such as network bandwidth limitations and fluctuating device thermal states. Although the evaluated algorithms all showed advantages and limitations, our findings demonstrate that Active Inference is a promising approach for ensuring SLO compliance in DCCS, offering lower memory usage, stable CPU utilization, and fast convergence.
\end{abstract}


\section{Introduction}

Over the decades, computing environments have evolved cyclically, shifting between centralized, decentralized, and distributed models based on technological advancements and organizational needs \cite{10207712,king1983centralized,donta2023exploring}. Recently, distributed environments, particularly those incorporating edge computing and IoT, have gained prominence due to their advantages, such as low latency and enhanced privacy \cite{10621659,beckman2020harnessing}. In this ongoing evolution, Distributed Computing Continuum Systems (DCCS) have emerged as a powerful approach, efficiently integrating multiple computational tiers into a cohesive ecosystem that ensures trade-off between cost, Quality of Service (QoS), and resource utilization at scale \cite{dustdar2022distributed}. In DCCS, tasks are allocated dynamically based on multiple criteria, including proximity, capacity, cost, and priority, enhancing real-time processing and minimizing latency. Unlike traditional edge computing, they offer fault tolerance by reallocating tasks to other available servers in case of device failure, ensuring uninterrupted computation \cite{donta2023governance, pujol2024causality}. Additionally, DCCS prioritize resource efficiency, enabling scalable and adaptive computing across the continuum. They maintain a high Quality of Experience (QoE) despite changing system requirements and environmental uncertainties by effectively managing resources \cite{casamayor2023fundamental, pujol2023edge}.

Simultaneously, DCCS are complex, open systems, vulnerable to workload spikes, evolving requirements, and dynamic infrastructure changes \cite{dustdar2022distributed}. These challenges necessitate adaptive capabilities to maintain optimal performance and resilience. However, anticipating all possible system configurations and environmental conditions is often impractical. Therefore, DCCS require pervasive intelligence across the entire continuum to ensure seamless integration, optimal performance, and robust management. This intelligence enables the system to respond to fluctuations in workload and changing conditions dynamically, ensuring reliability and efficiency in real-time processing and resource allocation \cite{donta2024human_based}. Thus, Service Level Objectives (SLOs) \cite{zhao2024SLOpt,sedlak2024diffusing,lin2024murmuration,casamayor2024deepslos} are introduced to provide a structured approach for monitoring, predicting, and managing or adapting system behavior across diverse computing environments. 
By establishing clear performance targets, SLOs enable adaptive mechanisms that respond dynamically to fluctuations in workload and changing conditions, ensuring that the system meets predefined performance standards. 

In the literature, various mechanisms have been explored for the dynamic adaptation of workloads through effective orchestration strategies \cite{kokkonen2022autonomy}. While these topics are gradually addressed by applying different machine learning (ML) mechanisms, e.g., \cite{sedlak_towards_2025,zhang_octopus_2023}, these works fail to provide a fundamental understanding of which ML techniques to apply to ensure SLOs. For example, with \textit{Octopus}~\cite{zhang_octopus_2023}, the authors created an SLO-aware inference scheduler based on Advantage Actor-Critic (A2C). For evaluation, the common scheme here is to use baselines designed for a different use case or have a completely different architecture. Within \textit{Octopus}, the open question is whether other ML techniques, e.g., Proximal Policy Optimization (PPO), would have performed superiorly. 
To provide a profound understanding of how different ML techniques rank for ensuring SLOs in a DCCS application, this paper provides benchmarks that target various aspects of the techniques. While there exist benchmarking solutions for pure Edge computing~\cite{cilic_performance_2023}, the evaluated solutions were not intended for the  DCCS. Also, runtime adaptations for stream processing~\cite{cardellini_runtime_2022} have a long history, but they are designed for static requirements, while SLOs, derived from a business context, are an evolving system property. 

Meanwhile, Active Inference (AIF)~\cite{parr2022active}---a concept from neuroscience---is gaining significant attention due to its ability to efficiently predict and adapt to changing conditions. AIF attempts to explain the behavior and learning of sentient creatures; to raise the level of intelligence in DCCS, AIF is also increasingly adopted in computer science. Recent literature \cite{sedlak2024equilibrium, sedlak2024active} has shown that AIF agents can effectively ensure SLO compliance, maintain high QoS and QoE, and continuously learn and adapt to dynamically changing environments and requirements. The promising results of AIF in DCCS inspire further exploration and also draw our curiosity to evaluate its performance against Reinforcement Learning (RL)-based algorithms (Deep Q-Network (DQN) \cite{mnih2015human}, A2C \cite{mnih2016asynchronous}, and PPO \cite{schulman2017proximal}), which have recently gained popularity and demonstrated significant benefits across various applications.

To the best of our knowledge, we are thus the first to provide a benchmarking solution for dynamic SLO compliance. During our study, we found that many existing works are conducted under simplified assumptions, lacking the complexity of real-world application scenarios. Hence, we simulate realistic video conferencing applications to rigorously test the aforementioned algorithms, ensuring a comprehensive evaluation of their performance and adaptability. In this context, our contributions are threefold, as outlined below:
\begin{enumerate}
    \item To evaluate algorithms in a realistic environment, we implement a custom DCCS use case that contains (i) an edge device running a video conferencing application and (ii) a WebSocket server streaming videos to the edge device. To ensure device SLOs, the server hosts an intelligent agent that optimizes the number of video streams, frames per second (FPS), and video resolution.
    
    \item We provide a benchmark for dynamic QoS and QoE fulfillment in DCCS. This simplifies the design choice for stakeholders by providing insights into the different capabilities of AIF and common RL algorithms.
    
    \item To further compare the algorithms' robustness against dynamic changes in environment or system requirements, we perform a series of experiments, where:

    \begin{enumerate}
        \item We introduce an instant distribution shift by significantly limiting network bandwidth.
        
        \item We introduce a gradual distribution shift by simulating an overheating device. This shows if algorithms can differentiate between dynamic system evolution and environmental noise.
        
        \item We change SLO thresholds to see whether algorithms can dynamically adapt to new objectives.
    \end{enumerate}
\end{enumerate}

The remaining sections of this paper are organized as follows: Section \ref{sec:priliminaries} provides an overview of AIF and common RL algorithms. Section \ref{sec:methodology} presents a detailed use case, SLO design, and algorithm implementation. In Section~\ref{section:experiments}, we provide a detailed discussion of the various criteria and scenarios used for evaluating the benchmarks, along with the experimental setup. Section \ref{sec:results} offers extensive results and discussions, along with a summary of limitations and potential extensions. Finally, we conclude the paper in Section \ref{sec:conclusions}.

\section{Related Work} \label{sec:priliminaries}

Although the merits of DCCS are openly discussed in recent research~\cite{donta2023exploring,cardellini_scalable_2025}, the heterogeneity and dynamism of DCCS are open challenges.
To enhance the rigor and reproducibility of our benchmarks, we compare AIF against three well-established RL techniques, which serve as baseline approaches. Each technique offers distinct advantages in ensuring SLO compliance. In this section, we provide a brief overview of how these methods were applied in the context of dynamic SLO management and analyze their respective strengths and limitations. To effectively highlight the differences between AIF and RL-based approaches, we categorize the three RL techniques---DQN, A2C, and PPO.

\subsection{Reinforcement Learning}
Dynamic processing environments often suffer from fluctuating workload patterns or multiple competing SLOs. To ensure SLO compliance under these circumstances, RL has been applied for proactive orchestration, e.g., using A2C to adjust to client pattern~\cite{qiu_firm_2020} or using DQN to find a trade-off between scaling actions~\cite{sedlak_towards_2025}. Particularly for autoscaling, DQN are applied by numerous researchers, e.g., for ensuring high utilization in the Cloud~\cite{yalles_riscless_2022} or adjusting the size of serverless containers~\cite{filinis_intent-driven_2024}. However, the authors chose their respective RL algorithms based on expert knowledge and experience, i.e., not supported by empirical evidence. 

Within the RL family, the three algorithms (i.e., DQN, A2C, and PPO) have characteristic strengths and weaknesses in terms of sample efficiency and stable convergence. 
As we will see later in the results, this proves critical at cold starts with few training samples or during distribution shifts.

\subsection{Active Inference}

Although AIF is not as widely applied for ensuring SLO compliance, it has found its way from neuroscience, over robotics, to computing systems~\cite{vyas_towards_2025}. In contrast to RL, the challenge is not to maximize the expected reward but to minimize \textit{free energy}, a measure of the uncertainty in the environment. More precisely, AIF agents must constantly balance between actions that improve its understanding of the environment and such that ensure high pragmatic value, i.e., fulfill SLOs. One option for creating a model of the environment is to train knowledge graphs from observations, as done in~\cite{sedlak2024equilibrium,sedlak2024adaptive}. While training these structures poses an overhead, they improve trustworthiness because the behavior of agents can be traced empirically.

However, to the best of our knowledge, there exist no scientific works that performed extensive evaluations between RL and AIF techniques, which again leaves the choice with the developer according to personal benefit. To support stakeholders in making this design choice, the benchmarks created in this paper will provide a profound idea of the advantages of each technique. In the following, we describe how these different algorithms are incorporated into our methodology.

\section{Methodology}\label{sec:methodology}
This section introduces an extensible benchmarking platform designed to ensure SLO compliance in dynamic DCCS environments. We begin by presenting a real-time video conferencing use case, incorporating a realistic environment setup. Then, we provide a comprehensive discussion on SLO composition, considering various quality metrics. Finally, we detail the implementation of key algorithms---AIF, DQN, A2C, and PPO---while highlighting their primary hyperparameters.

\subsection{Use Case}

We consider a real-time video communication service as a case study to compare the algorithms in realistic conditions. Our simulated environment has two components:
\begin{enumerate}
    \item \emph{Client:} This is a video-conferencing application that runs on an edge device (e.g., iPhone). By using the application, a user may join a conference with $N$ participants, where each participant provides a video stream. Each video stream is characterized by resolution and FPS. 

    \item \emph{Server:} Provides a configurable video stream to clients. To ensure high QoS and QoE for clients, the server hosts an intelligent agent that continuously learns an optimal policy (i.e., streaming configuration) through one of the compared algorithms.
\end{enumerate}

\subsubsection*{Streaming Process}

\begin{figure}[t]
    \centering
    \includegraphics[width=1\linewidth]{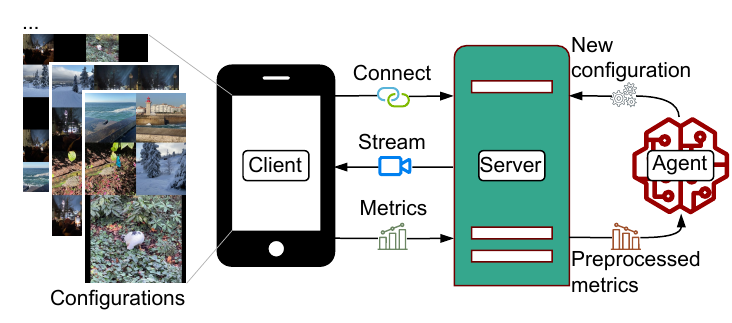}
    \caption{Overview of the Streaming Process}
    \label{fig:streaming_process}
\end{figure}

Initially, the client connects to the server, which is set up with a default configuration and SLOs (refer to Table~\ref{tab:exp_slos}).
The server begins to stream videos to the client, as visualized in Fig. \ref{fig:streaming_process}. While rendering the streams, the client locally collects performance metrics and transmits them to the server at fixed intervals.
The server then uses these metrics to train its agent and infer a client configuration that should improve SLO compliance.
The client is then instructed to operate with this new streaming configuration. This process is repeated throughout the entire lifetime of the client-server connection.

It is important to note that while local decision-making is preferred, we implement learning and inference on the server to accelerate simulation experiments through parallel execution in the cloud using pre-collected metrics and to enhance reproducibility. Additionally, server-side implementation allows us to leverage stable and reliable libraries.

\subsection{SLO Composition}

With the use case set, it remains to describe how the DCCS application will be monitored and configured. For this, we capture a set of metrics that give insights into the performance and efficiency of the streaming pipeline. During runtime, a set of SLOs must be fulfilled; in case they are violated, the server can act by changing the streaming configuration.

\subsubsection*{\textbf{Metrics}}
To quantify SLO compliance, train the agent, and infer the next system configurations, the client collects various performance metrics, including \emph{CPU usage} ($M_{CPU}$), \emph{memory usage} ($M_{mem}$), \emph{throughput} ($M_{tp}$), \emph{average latency} ($M_{lat}$), \emph{average render scale factor} ($M_{rs}$) and \emph{thermal state} ($M_{ts}$).

Consider a system with dynamically changing configurations, indexed by configuration timesteps $c\in\mathbb{N}^+$. At each configuration timestep $c$, there is a set of video streams indexed by $i\in\{1,\dots,N_c\}$. Each video stream $i$ emits video frames indexed by frame timesteps $t\in\mathbb{N}^+$, where each frame has a size (in bytes) denoted as $b_i(t)$. These frames may be captured at different real-world timestamps $\tau_i(t)$. Every second, for each configuration $c$ applied during that time interval, the client receives $T_c$ sets of video frames, represented as $\{F_i(t)|\forall i\}_{t=1}^{T_c}$. In practice, configurations do not change that frequently. Based on this setup, the \emph{average latency} is measured as shown in Eq.~(\ref{equation1_latency})
\begin{equation}\label{equation1_latency}
    M_{lat}=\alpha\sum_{t=2}^{T_c}\beta\sum_{i=1}^{N_c}(\tau_i(t)-\tau_i(t-1))
\end{equation}
\noindent where $\alpha=1\mathbin{/}(T_c-1)$ and $\beta=1\mathbin{/}N_c$. We calculate \emph{throughput} as the average amount of data received over the time period $T_c$, as shown in  Eq.~(\ref{equation_throughput})
\begin{equation}\label{equation_throughput}
    M_{tp}=\alpha\sum_{t=2}^{T_c}\beta\sum_{i=1}^{N_c}\frac{b_i(t)}{\tau_i(t)-\tau_i(t-1)}.
\end{equation}
%
The \emph{average render scale factor} is given by:
\begin{equation}
    M_{rs}=\alpha\sum_{t=2}^{T_c}\beta\sum_{i=1}^{N_c}\sqrt{\frac{W_i(t)\times H_i(t)}{w_i(t)\times h_i(t)}}
\end{equation}
\noindent where $w_i(t)$ and $h_i(t)$ are the pixel width and height of the video stream, and $W_i(t)$ and $H_i(t)$ represent the corresponding dimensions of the rendered area on the client device screen at timestep $t$.
The \emph{CPU usage} is calculated according to Eq.~(\ref{equation_CPU}).
\begin{equation}\label{equation_CPU}
    M_{CPU}=\alpha\sum_{t=2}^{T_c}\frac{U_{act}(t)}{U_{ref}}
\end{equation}
\noindent where $U_{act}(t)$ is the actual CPU usage at timestep $t$, and $U_{ref}$ is the expected maximum CPU usage, which we set to 200\%.
Similarly, \emph{memory usage} is computed as shown in Eq.(\ref{eqaution_Memory})
\begin{equation}\label{eqaution_Memory}
    M_{mem}=\alpha\sum_{t=2}^{T_c}\frac{R_{act}(t)}{R_{ref}}
\end{equation}
\noindent where $R_{act}(t)$ denotes actual memory usage at timestep $t$, and $R_{ref}$ the expected maximum memory usage, which we set to 200Mb.
Finally, the \emph{thermal state} of the device is determined as:
\begin{equation}
    M_{ts}=\max_{t\in\{2,\dots,T_c\}}\Theta(t)
\end{equation}
\noindent where $\Theta(t)\in\{0,\dots,3\}$ represents the device's thermal state at timestep $t$, with 0 indicating a \emph{nominal} state, 1 indicates a \emph{fair} state, 2 indicates a \emph{serious} state and 3 representing a \emph{critical} state.

To ensure high QoE and QoS, the intelligent agent is continuously learning system configurations that comply with the following SLOs.
For any metric $M_x$, where $x$ represents a placeholder for any metric type, variables $M_x^{max}$ and $M_x^{min}$ denote the upper and lower SLO thresholds, respectively.
For an overview of possible assignments, refer to Table~\ref{tab:exp_slos}.
\begin{enumerate}
    \item \emph{Average render scale factor ($M_{rs}^{max}$):} We aim to display streams of a sufficient resolution on a client device, ensuring that $M_{rs}\leq M_{rs}^{max}$. Maintaining $M_{rs}$ within this range enhances the streaming experience and ensures that video content does not appear blurred.

    \item \emph{Stream fulfillment ($M_{sf}^{min}$):} To fulfill its purpose, the video conferencing application should show a minimum number of streams from connected participants. 

    \item \emph{Average latency ($M_{lat}^{max}$):} To ensure high QoE, our services should provide a pleasant streaming latency for viewers. Therefore, we wish to maintain a sufficiently low latency, averaged over all received video streams.

    \item \emph{Throughput ($M_{tp}^{max}$):} High network usage drains the smartphone battery and puts excessive load on the server, and hence, may turn out costly. Therefore, we aim to constrain the device's throughput to a certain limit.

    \item \emph{Thermal state ($M_{ts}^{max}$):} Excessive resource usage and a hot environment may heat the device dangerously. Our application should respond to it by adjusting the system configuration to facilitate cooling.
\end{enumerate}

We calculate SLO-compliance level $S_x\in[0,1]$ for any given metric $M_x$ according to the following formula:
\begin{align} \label{eq:slo_calculation}
    S_{x}=\begin{cases}
        \min(1, M_{x}^{max}\mathbin{/}M_{x}), &\text{if $M_{x}^{max}$ is defined}\\
        \min(1, M_{x}\mathbin{/}M_{x}^{min}), &\text{if $M_{x}^{min}$ is defined}\\
    \end{cases}
\end{align}

Furthermore, we calculate the SLO-compliance levels for QoE and QoS SLOs according to Eq.~\eqref{eq:qoe_slo} and Eq.~\eqref{eq:qos_slo}. Then, we calculate the overall SLO compliance according to Eq. (\ref{eq:overall_slo}). Each SLO-compliance level is in the range $[0, 1]$, with 1 being ideal.
\begin{equation} \label{eq:qoe_slo}
    S_{QoE}=(S_{rs}+S_{sf})\mathbin{/}2
\end{equation}
\begin{equation} \label{eq:qos_slo}
    S_{QoS}=(S_{lat}+S_{tp}+S_{ts})\mathbin{/}3
\end{equation}
\begin{equation} \label{eq:overall_slo}
    S=(S_{QoE}+S_{QoS})\mathbin{/}2
\end{equation}

\subsubsection*{\textbf{Actions}}

To maximize SLO compliance, the server can take action, namely, change the streaming configuration. The respective policy---which configuration to choose to optimize SLOs---is learned over time. The system configuration itself is characterized by the following streaming parameters:
\begin{enumerate}
    \item \emph{Number of streams:} The number of streams the client receives and subsequently renders on the screen.

    \item \emph{Resolution:} Pixel dimensions of video streams.

    \item \emph{FPS:} Frame rate of video streams.
\end{enumerate}

We constrain the parameters to the following possible values: $\mathcal{A}_{streams}=\{1,2,5,10,15,20\}$, $\mathcal{A}_{resolution}=\{180,360,720\}$ (equal widths and heights) and $\mathcal{A}_{fps}=\{5,10,15,20,25,30\}$. This forms an action (configuration) space $\mathcal{A}=\mathcal{A}_{streams}\times\mathcal{A}_{resolution}\times\mathcal{A}_{fps}$ of size 108.

\subsection{Algorithm Implementations} 

\begin{table}[t]
    \caption{Hyperparameters}
    \centering
    \resizebox{\columnwidth}{!}{%
    \begin{tabular}{|l|c|c|c|c|}
        \hline
        \textbf{Hyperparameter} & AIF & DQN & A2C & PPO\\
        \hline
        surprise\_threshold\_factor & 2.0 & -- & -- & --\\
        weight\_of\_past\_data & 0.6 & -- & -- & --\\
        initial\_additional\_surprise & 1.0 & -- & -- & --\\
        graph\_max\_indegree & 8 & -- & -- & --\\
        hill\_climb\_epsilon & 1.0 & -- & -- & --\\
        input\_size & 32 & 1 & 1 & 1\\
        batch\_size & 32 & 128 & 64 & 128\\
        learning\_rate & -- & $10^{-4}$ & $10^{-4}$ & $10^{-4}$\\
        exploration\_initial\_eps & -- & 1.0 & -- & --\\
        exploration\_final\_eps & -- & 0.05 & -- & --\\
        exploration\_fraction & -- & 0.1 & -- & --\\
        train\_freq & -- & 4 & -- & --\\
        grad\_steps & -- & 4 & -- & --\\
        target\_update\_interval & -- & 10000 & -- & --\\
        neurons & -- & [128, 128] & [128, 128] & [64, 64]\\
        gamma & -- & 0.99 & 0.99 & 0.99\\
        gae\_lambda & -- & -- & 0.9 & 0.95\\
        vf\_coef & -- & -- & 0.75 & 0.25\\
        ent\_coef & -- & -- & 0.01 & 0.01\\
        normalize\_advantage & -- & -- & TRUE & TRUE\\
        n\_steps & -- & -- & -- & 1280\\
        n\_epochs & -- & -- & -- & 10\\
        clip\_range & -- & -- & -- & 0.2\\
        \hline
    \end{tabular}
    }
    \label{tab:hyperparams}
\end{table}

In this subsection, we discuss the implementations of the evaluated algorithms. To achieve stable performance and high SLO compliance, we optimize their key hyperparameters using a grid search approach; the selected values are summarized in Table~\ref{tab:hyperparams}. When multiple hyperparameter assignments showed identical performance, we preferred the most efficient assignment in terms of CPU consumption time and memory usage.

\subsubsection{Active Inference} \label{section:aif}
Our implementation of the AIF agent is based on \cite{sedlak2024equilibrium} with several adjustments, reflecting the increased complexity of our evaluation environment and facilitating fair comparison with RL methods. Specifically, we introduce changes to the metric pre-processing and computation and interpolation of a \emph{pragmatic value} (\emph{pv}) and \emph{risk assigned} (\emph{ra}). These values, together with \emph{information gain} (\emph{ig}), serve as criteria for selecting the next system configuration.

In the original approach, all metrics are discretized, with SLO-related metrics being converted to binary values, according to Eq. (\ref{eq:binary_metric}).
\begin{equation} \label{eq:binary_metric}
    f(M_x)=\begin{cases}
        1 &\text{if }S_x=1\\
        0 &\text{otherwise}
    \end{cases}
\end{equation}

These discrete and binary values are used to learn the structure and parameters of a generative model used by AIF. Consequently, \emph{pv} and \emph{ra} are computed as a joint probability of SLO compliance with respect to all QoE and QoS SLOs, accordingly. {For example, let $A$ be a random variable taking values from the action space $\mathcal{A}$.} Given a configuration $a\in\mathcal{A}$, \emph{pv} and \emph{ra} are computed according to Eq. (\ref{eq:pv_1}) and Eq. (\ref{eq:ra_1}), respectively.
\begin{equation} \label{eq:pv_1}
    pv_a=P(S_{QoE}=1|A=a)
\end{equation}
\begin{equation} \label{eq:ra_1}
    ra_a=P(S_{QoS}=1|A=a)
\end{equation}

Our baseline approach has certain limitations; for example, it does not consider the partial SLO compliance in system configurations.
Each SLO variable is in the continuous range $[0,1]$ (refer to Eq. (\ref{eq:slo_calculation})), which is not captured by binary variables. Further, there may be a high probability that at least one of the QoS and QoE SLOs is fulfilled, but this is ignored when considering joint distribution only. These assumptions lead the algorithm to treat both suboptimal and partially compliant configurations equivalently. While this may suffice for straightforward scenarios, it introduces considerable bias in more complex environments, particularly when achieving full SLO compliance (i.e., $S = 1$) is infeasible. Therefore, we introduce several modifications to the AIF implementation to enhance its competitiveness with RL, enabling it to effectively distinguish between varying levels of SLO compliance through reward mechanisms.

First, we continue discretizing non-SLO metrics, namely, \emph{CPU time} and \emph{memory usage}. We partition these variables into a set of ordered, non-overlapping intervals $\{[0,0.2],(0.2,0.4],(0.4,0.6],(0.6,0.8],(0.8,+\infty)\}$ and assign each observation a discrete value from $\{1,\dots,5\}$ based on an interval it falls in. However, we also do a similar procedure with SLO-compliance levels, but with intervals $\{[0,0.2),[0.2,0.4),[0.4,0.6),[0.6,0.8),[0.8,1),\{1\}\}$ and labels $\{0,\dots,5\}$, respectively. In this case, we define this mapping as the discretization operator $\mathcal{D}:\mathbb{R}\rightarrow\mathbb{N}$. This formulates \emph{pv} and \emph{ra} more precisely and approximates expectations over $S_{QoE}$ and $S_{QoS}$. We calculate the true expected SLO compliance given a configuration $a\in\mathcal{A}$ through
\begin{equation} \label{eq:true_expected_slo}
    f_x(a) = \mathbb{E}[S_{x}|A=a]=\frac{1}{|\mathcal{S}_x|}\sum_{S_i\in \mathcal{S}_x}\int_0^1s_ip(s_i|a)ds_i
\end{equation}
\noindent where for \emph{pv} we use $f_{QoE}(a)$ with $\mathcal{S}_{QoE}=\{S_{sf},S_{rs}\}$ and for \emph{ra} $f_{QoS}(a)$ with $\mathcal{S}_{QoS}=\{S_{lat},S_{tp},S_{ts}\}$, respectively. However, as we discretize SLOs, calculating true expectation is impossible; therefore, we approximate this function as shown in Eq. (\ref{eq:approx_expected_slo}):
\begin{equation} \label{eq:approx_expected_slo}
    \tilde{f}_x(a) =\frac{1}{|\mathcal{S}_x|}\sum_{S_i\in\mathcal{S}_x}\sum_{\tilde{s}\in \mathcal{D}(S_i)}\mu(\tilde{s})P(\tilde{s}|a)
\end{equation}
\noindent where $\mu(\tilde{s})=\min(1, 0.2\tilde{s}+0.1)$ approximates a true SLO-compliance value. This approach allows AIF to differentiate more precisely between partially SLO-complying system configurations, which is critical for fair comparison with RL.

Additionally, in Sedlak \textit{et al.}~\cite{sedlak2024equilibrium}, the configuration space is two-dimensional, and they artificially increase initial \emph{ig} values for key configurations to facilitate the interpolation of \emph{pv} and \emph{ra} matrices for unexplored parameters. In our work, the configuration space is three-dimensional, therefore, we use tensors. Similarly, we increase initial \emph{ig} values for 8 configurations, located in the corners of this three-dimensional space. These represent all possible combinations of $\{1,20\}$ streams, $\{180,720\}$ resolutions and $\{5,30\}$ FPS values.

\subsubsection{Reinforcement Learning}

To oppose AIF, we benchmark it with three well-established RL algorithms, namely, DQN, A2C, and PPO. We use the implementations from Stable Baselines3 (SB3)~\cite{stable-baselines3}, which are well-documented and standardized for this purpose.

We standardize continuous metrics (\emph{CPU usage}, \emph{memory usage}, \emph{throughput}, \emph{average latency} and \emph{average render scale factor}) based on pre-collected metrics data (see Section \ref{section:experiments}), and one-hot encode a categorical metric (\emph{thermal state}) and system configuration parameters. We stack together the resulting features into a vector of size 24. This vector corresponds to an observation processed by DQN, A2C, and PPO algorithms. For each observation, we compute a reward $\mathcal{R}$, corresponding to a SLO-compliance level for this observation as in Eq. (\ref{eq:overall_slo}) (i.e., $\mathcal{R}\equiv S$). With such observations and rewards, RL algorithms have access to all available information and can differentiate between partially complying configurations, which enables fair comparison with AIF.

\section{Experimental Design and Test Cases} \label{section:experiments}

We conduct a series of experiments to evaluate the performance and efficiency of the benchmark algorithms in terms of SLO compliance within dynamic and uncertain DCCS environments. 
{Initially, we pre-collect a dataset (refer to Data Availability) containing various system performance metrics gathered from controlled experimental conditions prior to simulation. This static dataset serves as the foundation for accurately modeling the dynamics of our environment.}
We sequentially accumulate metrics for each possible system configuration over several minutes in real time. The dataset used in experiments encompasses 512 records of metrics per configuration. During experiments, when an agent selects a new configuration, we sample a batch of corresponding metrics from the dataset as if they were collected in real time. This approach makes training and evaluation incomparably faster and facilitates the reproducibility of results as we publish our data. Our experiments are evaluated under both certain and uncertain conditions, including basic preferences, instantaneous and gradual distribution shifts, and changing SLO requirements, which are detailed below.

\begin{table}[t]
    \centering
    \caption{Experiment SLOs}
    \resizebox{\columnwidth}{!}{%
    \begin{tabular}{|l|l|l|l|l|l|}
    \hline
    \textbf{Experiment} & $M_{tp}^{max}$ & $M_{lat}^{max}$ & $M_{sf}^{min}$ & $M_{rs}^{max}$ & $M_{ts}^{max}$\\
    \hline
    Basic & 10 Mb/s & \nicefrac{1}{15} s & 5 & 1.6 & 1\\
    Instant Shift & 10 Mb/s & \nicefrac{1}{15} s & 5 & 1.6 & 1\\
    Gradual Shift \#1 & 10 Mb/s & \nicefrac{1}{15} s & 5 & 1.6 & 1\\
    Gradual Shift \#2 & 10 Mb/s & \nicefrac{1}{15} s & 5 & 1.6 & 1\\
    Changing SLOs \#1 & 256 Kb/s & \nicefrac{1}{30} s & 20 & 0.25 & 1\\
    Changing SLOs \#2 & 5 Mb/s & \nicefrac{1}{15} s & 10 & 1.0 & 1\\
    \hline
    \end{tabular}
    }
    \label{tab:exp_slos}
\end{table}

\subsubsection{Basic Performance and Efficiency Evaluation} \label{section:exp_basic_performance}
Initially, we perform a basic experiment where the algorithms operate in regular environment conditions and are required to meet the SLOs in Table~\ref{tab:exp_slos}. During this experiment, we run algorithms for $1.28\times10^6$ environment steps (RL processes individual observations, while AIF batches of 32), each 6,400 steps performing evaluation. Evaluation involves executing inference deterministically, without learning in a separate copy of the environment. The evaluation sequence is repeated eight times at once, each time for 640 steps, starting from one of eight corner positions in our action space, $\mathcal{A}$. This approach allows us to summarize the performance of the algorithms with mean and standard deviation across various starting points. In this experiment, we evaluate the performance and efficiency of the algorithms by measuring SLO compliance, CPU time, and memory usage.

\subsubsection{Instant Distribution Shift} \label{section:instance_dist_shift}
DCCS are subject to unexpected network spikes, dynamically changing topology, and various failures. Hence, it is critical that algorithms can detect and handle such challenges. To assess the adaptable capabilities of the algorithms, we conduct this experiment where a client is suddenly facing a significantly reduced network bandwidth (1 Mb/s). To simulate this, we implement this limitation in our video streaming server and pre-collect the respective metrics. Then, we proceed with the training and evaluation of the pre-trained models (building on the basic preferences and performance metrics discussed in the previous subsection) using this new dataset while maintaining the same SLOs as outlined in Table \ref{tab:exp_slos}.

\subsubsection{Gradual Distribution Shift} \label{section:grad_dist_shift}
Often, environmental dynamics may evolve more gradually. In such circumstances, an intelligent algorithm within DCCS should capture the trajectory of system metrics and act accordingly. This proactive property---crucial for DCCS \cite{dustdar2022distributed,morichetta2021roadmap,pujol2023edge}---is inspected in this experiment. 
Suppose a user is attending an online meeting via a mobile device and suddenly they expose the device to an additional workload. This raises the device temperature, potentially to a point that damages the device. To facilitate cooling, applications should reduce their resource consumption. 

To simulate device heating, at each environment step, we calculate a \emph{target temperature} $T^*$, which represents the temperature to which the device tends to heat up. We calculate it as a function of throughput (as shown in Eq.~(\ref{eq:network_t})), as in our case, it is a straightforward proxy to network, CPU, and other resource utilization:
\begin{equation} \label{eq:network_t}
    T^*=\min\left[1,\kappa\times \exp\left(\frac{\lambda M_{tp}}{1024\times1024}\right)\right]
\end{equation}
\noindent where the coefficients $\kappa=0.364$ and $\lambda=0.05$ were chosen to decrease the number of optimal configurations yet remain feasible for full SLO compliance. Then, we plug this into a Newton's law of cooling equation \cite{winterton1999newton} to calculate the temperature at the next environment state $T_{t+1}$:
\begin{equation}
    T_{t+1}=T^*+(T_t-T^*)\times \exp(-k)
\end{equation}
\noindent where $k$ represents the cooling constant, and we arbitrarily set its value to 0.03 or 0.07 to evaluate the behavior of algorithms under different temperature changing speeds, i.e., we conduct this experiment twice with different $k$. Finally, we linearly map $T_{t+1}$ to a discrete value $\Theta\in \{0,\dots,3\}$ to represent the device's thermal state. We do this to make the simulation closer to reality, as iOS exposes temperature as a similar discrete value.

Similar to the previous experiment, we continue running the simulation on the pre-trained models discussed in Subsection \ref{section:exp_basic_performance}.  Thus, we can simulate dynamically heating/cooling the device and introduce strong temporal dependence into our system to evaluate the proactivity of the algorithms.

\subsubsection{Changing SLOs} \label{section:exp_changing_slos}
In DCCS, devices may be exposed to changing requirements, which require the intelligent agents to adapt according to the circumstances. To evaluate this, we do exactly that---dynamically change SLOs. We conduct two similar experiments in which we modify the SLOs such that full compliance becomes unattainable. The distinction between the two experiments lies in the feasibility of the objectives: one experiment features objectives that are less feasible than the other (see Table \ref{tab:exp_slos}).
Similarly to the previous two experiment types, we continue this evaluation on the pre-trained models.

\subsection{Execution Setup}

We generate metrics datasets for experiments using an iPhone 16 simulator with iOS 18.2, available in Xcode IDE (16.2), installed on MacBook M2 Pro with macOS 15.0.1. The application runs in the foreground in portrait device orientation, as shown in Fig.~\ref{fig:streaming_process}. We execute hyperparameter tuning and the experiments on a machine with two 32 core Intel(R) Xeon(R) Gold 8358 CPU @ 2.6GHz with 512 GiB RAM.

\section{Results and Discussion}\label{sec:results}

To compare the SLO compliance of RL and AIF algorithms, we ran experiments simultaneously and collected the respective metrics. Recall that RL algorithms operate on individual observations, whereas AIF requires batches of 32. Hence, to present the results of the uniform scale, we compute averages of subsequences of 32 metrics for each RL algorithm. 
To cover distribution shifts during runtime, we decided to learn and improve the streaming configuration continuously; this potentially presents an overhead, so we monitor CPU and memory utilization. 
However, we present SLO-compliance metrics from evaluation to avoid noise introduced by learning and exploration. The curves and transparent regions in SLO-compliance charts represent means and standard deviations over batches of 32 observations from eight evaluation sequences performed.
Means are computed according to Eq. (\ref{eq:result_mean}):
\begin{equation} \label{eq:result_mean}
    \mu = \frac{1}{T\times N}\sum_{t=1}^T\sum_{i=1}^NS_i(t)
\end{equation}
\noindent our experiments assume $N=8$ is a number of evaluation sequences, $T=32$ is a number of environment steps in a batch and $S_i(t)$ is a SLO-compliance level for the $t$-th step of $i$-th evaluation.
In turn, standard deviations are computed according to the law of total variance:
\begin{align}
    \sigma &=\sqrt{\mathbb{E}[\operatorname{Var}(S|T)]+\operatorname{Var}(\mathbb{E}[S|T])}\\\nonumber
    &=\left(\frac{1}{TN}\sum_{t=1}^T\sum_{i=1}^N\left[S_i(t)-\frac{1}{N}\sum_{j=1}^{N}S_j(t)\right]^2\right.\\\nonumber
    &+ \left.\frac{1}{T}\sum_{t=1}^T\left[\frac{1}{N}\sum_{i=1}^NS_i(t)-\frac{1}{TN}\sum_{t=1}^T\sum_{i=1}^NS_i(t)\right]^2\right)^{\nicefrac{1}{2}}.
\end{align}
For visual interpretability, we smooth the curves by averaging over the last 15 batches. Additionally, since full SLO compliance is infeasible in some experiments, we plot their lines denoted by \emph{"Exp."} that represent an average SLO compliance of the most optimal configuration based on pre-collected metrics used for the corresponding experiment.

\subsection{Basic Preferences}

\begin{figure}[b]
    \centering
    \includegraphics[width=1\linewidth]{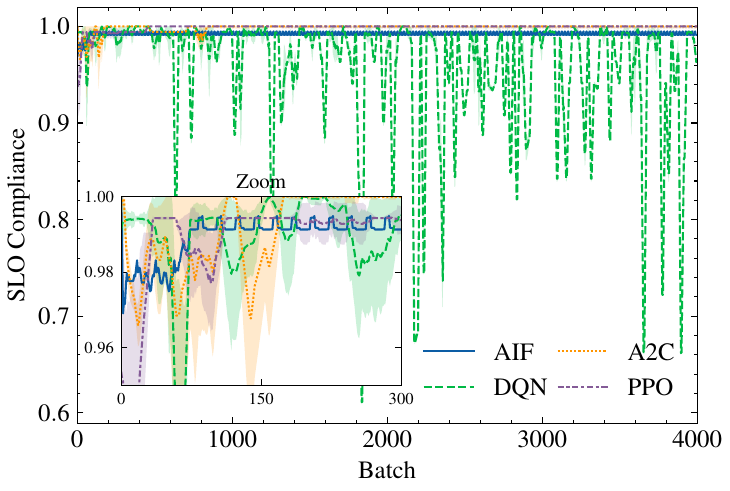}
    \caption{SLO Compliance during Basic Performance Evaluation}
    \label{fig:result_exp_basic}
\end{figure}

Fig. \ref{fig:result_exp_basic} shows that all four algorithms were able to achieve decently high SLO fulfillment rates under certain conditions. Specifically, AIF showed remarkable sample efficiency by converging significantly faster than other algorithms. However, its solution is mildly suboptimal, which may be partially attributed to environmental noise. In their turn, PPO and A2C were able to converge to an optimal configuration but required multiples of AIF training time, especially A2C. 

On the contrary, DQN showed the lowest SLO compliance and high instability. Although the initial evaluation cycles showed high performance, it progressively declined over time until around batch 2300, where improvement was observed.

Overall, the results suggest that in the base case AIF, PPO and A2C perform similarly, whereas DQN is distinguished by its instability, sample-inefficiency, and subpar performance.

\begin{figure}[t]
    \centering
    \subfloat[CPU time]{
    \includegraphics[width=0.497\columnwidth-0.5\columnsep]{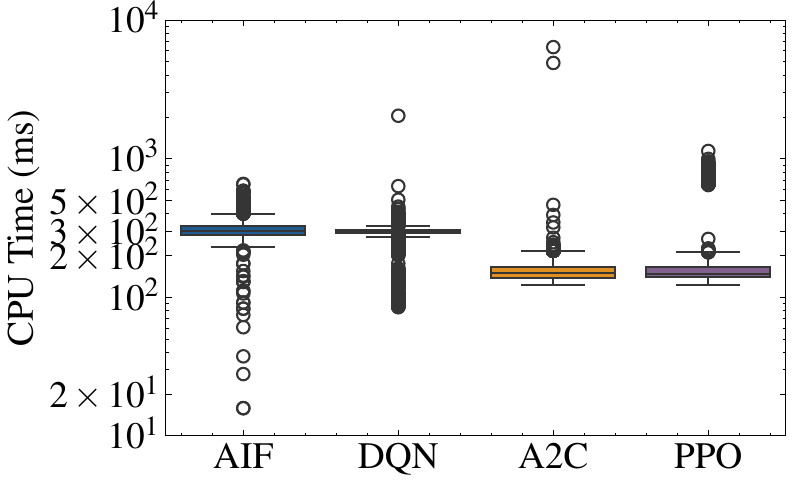}
    \label{fig:result_cpu}
    }
    \raisebox{-0.6em}{
    \subfloat[Memory usage]{
    \includegraphics[width=0.503\columnwidth-0.5\columnsep]{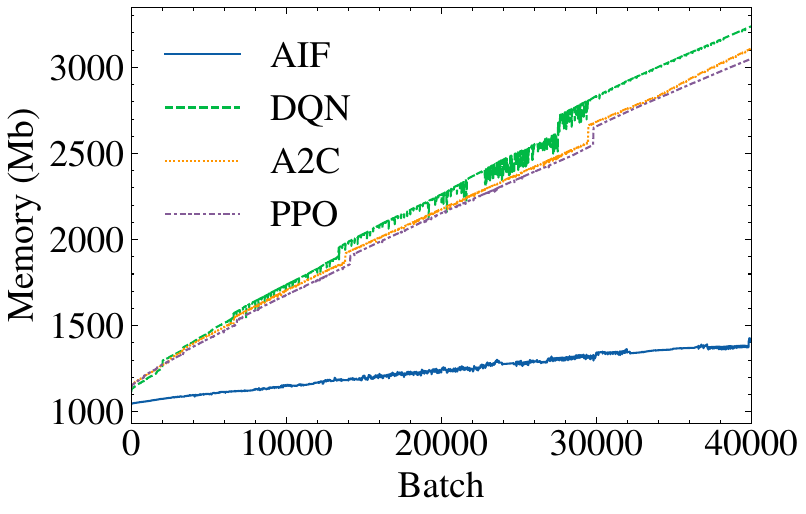}
    \label{fig:result_memory}
    }
    }
    \caption{Resource Utilization during Training}
    \label{fig:result_efficiency}
\end{figure}

In terms of efficiency, Fig. \ref{fig:result_cpu} demonstrates that AIF requires approximately equivalent CPU time as DQN on average, i.e., 304ms vs. 289ms, respectively. However, there are occurrences where DQN drastically exceeds this value, reaching up to 2.04s per batch. In contrast, AIF demonstrates numerous cases where it uses substantially less CPU time, up to 15ms.

A2C and PPO utilize approximately 2.3--2.5 times less CPU time than DQN and AIF on average, i.e., 121ms and 122ms, accordingly. However, they present considerably more outliers, reaching up to 6.39s for A2C and 1.13s for PPO per batch. Although occasional violations of our CPU utilization expectations in our setup are negligible, they can be critical in some applications, and this should be considered when choosing the right algorithm.

Overall, it is clear that the algorithms are comparable in terms of CPU utilization, with A2C and PPO being more efficient on average, though they occasionally introduce significant overhead. In contrast, AIF uses surprisingly less CPU time for some batches, demonstrating a more efficient use of resources.

Regarding memory utilization, Fig. \ref{fig:result_memory} suggests that all RL algorithms exhibit similar memory utilization dynamics, close to linear, with memory for DQN growing slightly faster than others. In contrast, AIF showcases a vastly lower memory footprint, which is rather logarithmic.
The memory usage of the AIF process increases by only 372 MB from the beginning to the end of training, whereas the three RL methods exhibit a significantly higher increase of 2.12 GB, 1.86 GB, and 1.90 GB for DQN, A2C, and PPO, respectively. Furthermore, by the end of training, the AIF process requires approximately 2.1--2.3 times less memory than RL methods. This makes AIF more advantageous for deploying to edge devices that have a tightly limited memory capacity.

\subsection{Distribution Shifts}
\subsubsection{Instant distribution shift}
\begin{figure}[t]
    \centering
    \includegraphics[width=1\linewidth]{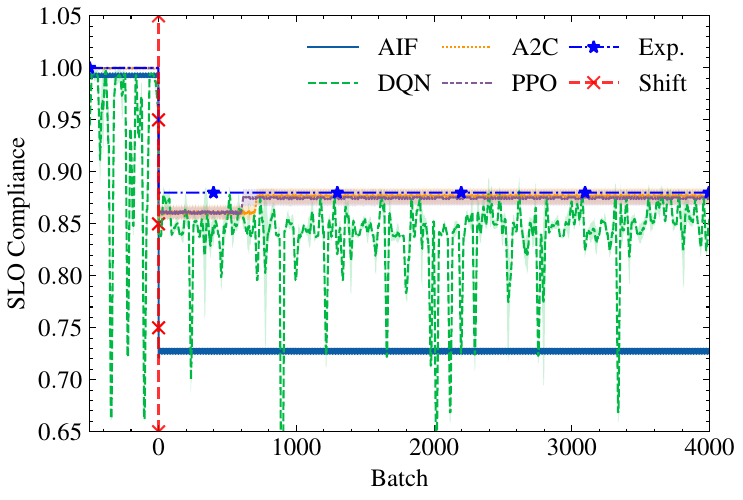}
    \caption{SLO Compliance after an Instant Distribution Shift}
    \label{fig:result_exp_rate_limit}
\end{figure}

\begin{figure*}[t]
    \centering
    \subfloat[large buffer, $h=2$ (default)]{
    \includegraphics[width=0.5\columnwidth-0.5\columnsep]{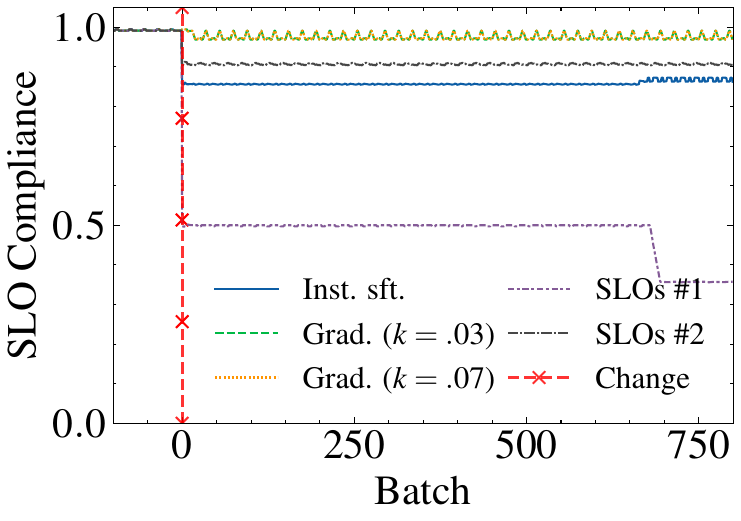}
    \label{fig:result_aif_eval_h2}
    }
    \subfloat[large buffer, $h=0$]{
    \includegraphics[width=0.5\columnwidth-0.5\columnsep]{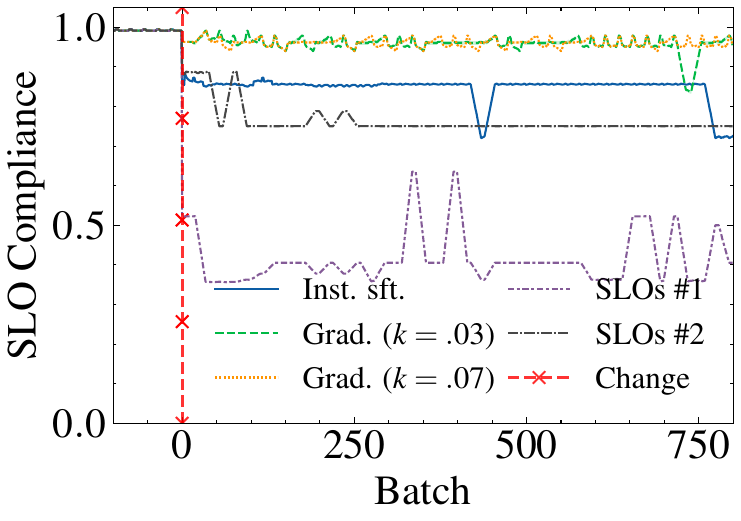}
    \label{fig:result_aif_eval_h0}
    }
    \subfloat[small buffer, $h=2$]{
    \includegraphics[width=0.5\columnwidth-0.5\columnsep]{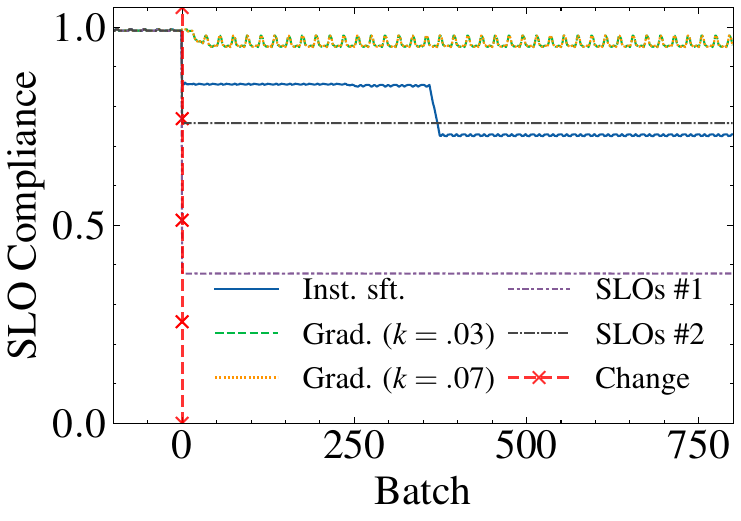}
    \label{fig:result_aif_eval_h2_sb}
    }
    \subfloat[small buffer, $h=0$]{
    \includegraphics[width=0.5\columnwidth-0.5\columnsep]{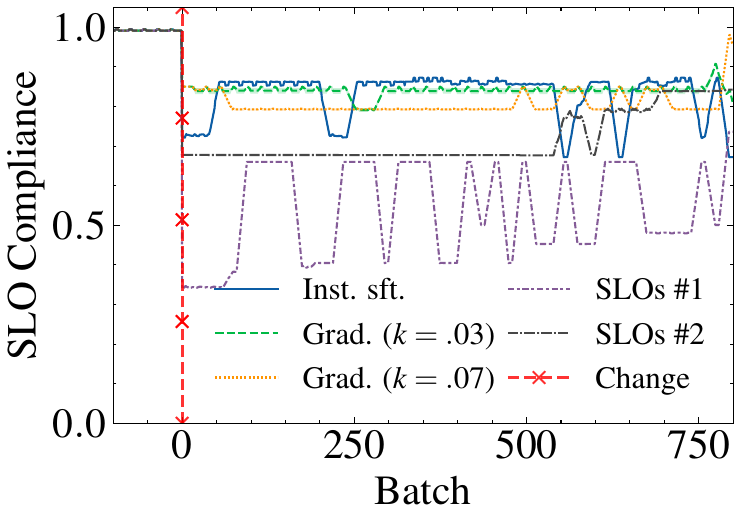}
    \label{fig:result_aif_eval_h0_sb}
    }
    \caption{SLO Compliance of AIF in each Experiment under Different Structure Learning Conditions (pre-trained on only 128,000 observations)}
\end{figure*}

{Fig. \ref{fig:result_exp_rate_limit} shows that PPO and A2C successfully detected the instant distribution shift and acted accordingly to reach near optimal SLO-compliance level. A2C converged slower but to a slightly better configuration.} Conversely, DQN demonstrates a decline in performance early on but continues to improve after batch 2000, eventually reaching SLO compliance close to PPO.
Surprisingly, AIF performs poorly. According to our investigation, this is caused by an optimization employed to reduce its computational overhead by limiting executions of structure and parameter learning of the generative model in \cite{sedlak2024equilibrium}, the model's parameters are re-learned only if $\Im_c>\tilde{\Im}_{10}$, where $\Im_c$ is the surprise caused by the current batch of data and $\tilde{\Im}_{10}$ is the median surprise over the last 10 batches; meanwhile, the model's structure is re-learned if $\Im_c>\tilde{\Im}_{10}\times h$, where $h$ is some factor. While these constraints significantly accelerate execution in the long run, they may also cause prolonged stagnation in a suboptimal configuration---one that is sufficiently surprising to be explored, yet not surprising enough to trigger model updates. 

Another issue with the AIF implementation is the incapability to forget outdated experiences. Fig. \ref{fig:result_aif_eval_h2} shows that AIF converged to a more optimal configuration after pre-training on 128,000 instead of 1,280,000 observations. Simultaneously, Fig. \ref{fig:result_aif_eval_h0} demonstrates AIF behavior when re-learning the model structure every batch. After the instant distribution shift, there are more fluctuations without performance improvements. This suggests that mixing new data distribution with the outdated one is not an effective strategy. Naively limiting the observation buffer does not contribute to improving results in this case (Fig. \ref{fig:result_aif_eval_h2_sb} and Fig \ref{fig:result_aif_eval_h0_sb}).

\subsubsection{Gradual distribution shifts}
\begin{figure}[t]
    \centering
    \subfloat[$k=0.03$]{
    \includegraphics[width=0.5\columnwidth-0.5\columnsep]{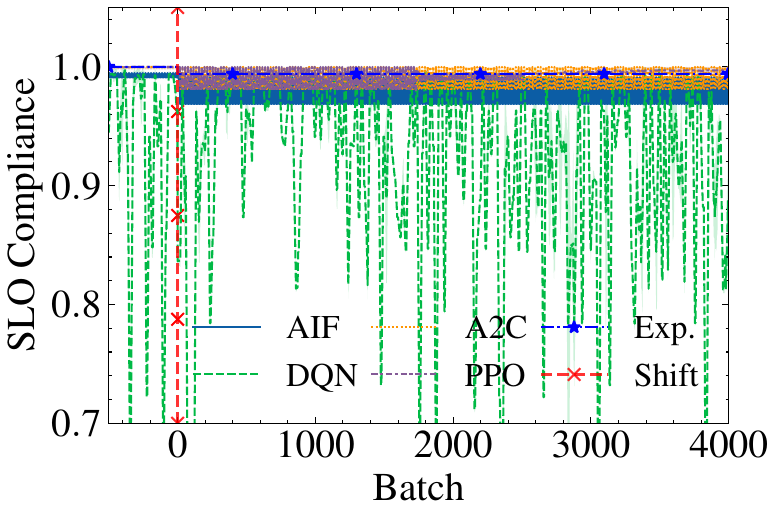}
    \label{fig:result_exp_thermal_03}
    }
    \subfloat[$k=0.07$]{
    \includegraphics[width=0.5\columnwidth-0.5\columnsep]{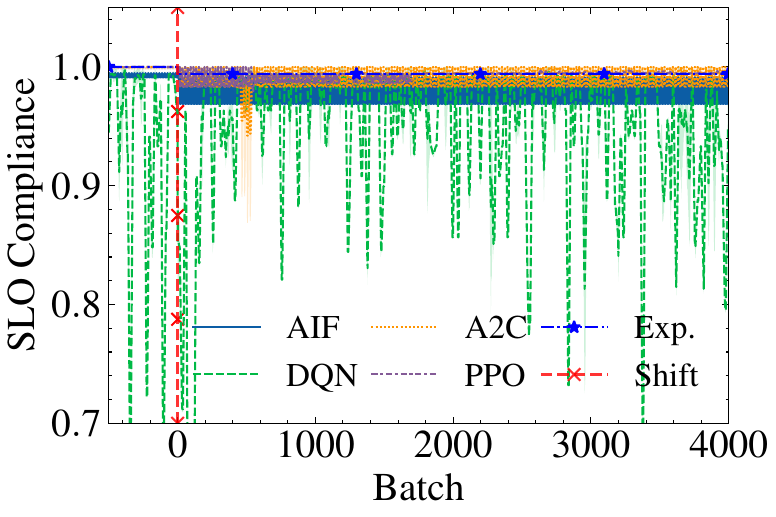}
    \label{fig:result_exp_thermal_07}
    }
    \caption{SLO Compliance under Changing Thermal States}\label{fig:result_exp_thermal}
\end{figure}

{In this scenario, Fig. \ref{fig:result_exp_thermal} shows that most algorithms struggle to effectively capture the pattern of thermal state variations. Increasing the parameter $k$, which corresponds to a faster update of thermal state, allows RL algorithms to consider a greater number of temperature changing cycles. Consequently, we expected higher $k$ values to improve stability and performance. As anticipated, DQN demonstrates better performance and greater stability when $k=0.03$ (Fig.~\ref{fig:result_exp_thermal_03}) compared to $k=0.07$ (Fig.~\ref{fig:result_exp_thermal_07}), though its overall performance remains suboptimal. PPO successfully adapts to different settings but exhibits a more gradual adaptation at $k=0.07$. Contrary to expectations, A2C displays increased fluctuations following a constant pattern, suggesting its inability to detect temperature changing trends. This behavior implies that dynamically changing discrete temperature poses a challenge for A2C, and it is worth exploring alternative techniques, such as utilizing continuous temperature. However, the reason for seemingly the same problem for AIF is different---the current implementation lacks capabilities to model relationships between consecutive states.}

As a result, it cannot effectively predict the long-term consequences of actions. This is also the reason why Fig. \ref{fig:result_aif_eval_h2}–\ref{fig:result_aif_eval_h2_sb} show very similar curves for experiments with temperature---AIF's generative model simply cannot capture dependencies of temperature on other variables, so, from AIF standpoint the data distribution did not change, with exception that fulfillment of the thermal state SLO became virtually random.
Therefore, to be generalizable to such non-stationary systems, AIF implementations should have the capacity to model transitions between states and plan series of actions to achieve long-term SLO compliance.

\subsection{Changing SLOs}

\begin{figure}[t]
    \centering
    \subfloat[Case \#1]{
    \includegraphics[width=0.5\columnwidth-0.5\columnsep]{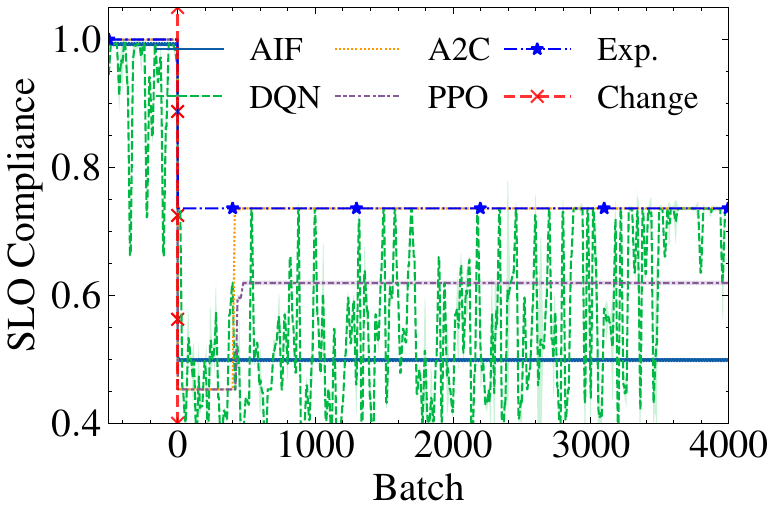}
    \label{fig:result_exp_slos_hard}
    }
    \subfloat[Case \#2]{\includegraphics[width=0.5\columnwidth-0.5\columnsep]{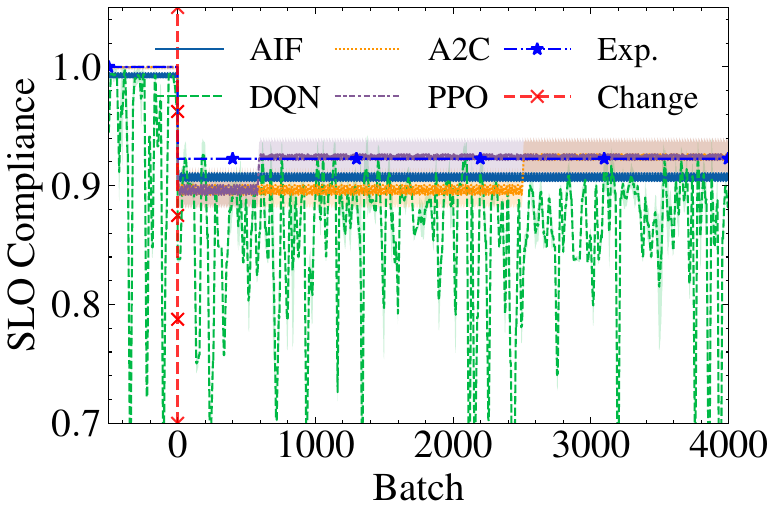}
    \label{fig:result_exp_slos_easy}
    }
    \caption{SLO Compliance after Changing SLOs}
\end{figure}

Similarly to the case of instant distribution shift (Fig. \ref{fig:result_exp_rate_limit}), in both cases of changing SLOs (Fig. \ref{fig:result_exp_slos_hard} and Fig.~\ref{fig:result_exp_slos_easy}) AIF suffers the same issue---getting stuck with a suboptimal configuration due to experiencing insufficient surprise to trigger model updates. However, relieving the limitations of model update frequencies shows even less stable and performant results (Fig. \ref{fig:result_aif_eval_h2}), which again prompts the need for exploration of more advanced techniques for handling distribution shifts. Although, Fig. \ref{fig:result_aif_eval_h2_sb} shows some improvement (even with less stability) in the case \#1 of SLO change, this correlates with other observations in Fig. \ref{fig:result_aif_eval_h2_sb}, where we see greater variance than in other figures, which stems from both, re-learning model every batch and having a small buffer with observations. Nonetheless, AIF can maintain a high SLO-compliance level in case \#2 and, with greater stability, {surpass DQN and even PPO and A2C until around the 600th and 2500th batch, respectively.}

DQN remains consistent with previous experiments, exhibiting high variance, but simultaneously, its performance slowly improves, and it approaches near optimal SLO compliance. 
{Interestingly, both PPO and A2C initially remain on the same SLO-compliance level and eventually can detect SLO change. 
In a simpler case (\#2), both algorithms converge to a near optimal configuration, and PPO converges significantly faster.
However, in case \#1 A2C converges to an optimal configuration, while PPO improves insufficiently. However, in case \#2, both converge to a near optimal configuration, and PPO converges significantly faster. This may stem from a higher stability of the PPO algorithm, influenced by methods like gradient clipping and the Kullback-Leibler divergence term, which hinders its exploration. Significant objective shifts could cause larger gradient updates, which fostered exploration in A2C, but in PPO, these gradients were clipped, and drastic policy changes were penalized.}

\subsection{Limitations and Future Work}

Based on our observations, AIF remains a promising approach in the context of DCCS, demonstrating lower memory footprint than other algorithms, fair and predictable CPU utilization, and fast convergence to near optimal SLO-compliance level in our standard scenario. Although PPO and A2C algorithms achieved high scores and stability in many cases, AIF is also inherently explainable and, via the use of Markov blankets, allows us to infer system configurations, discarding irrelevant factors and thus accelerating inference speed. This makes it attractive for embedding into highly complex systems that require dependability. With that said, we identify several issues with the current AIF implementation for DCCS that should be addressed in future work:

\begin{enumerate}
    \item AIF should capture relationships between consecutive observations and plan series of actions to predict effectively optimal system configurations in environments where observations exhibit interdependencies.

    \item AIF should more effectively balance updating a generative model and minimizing computational latency from frequent learning to avoid hindering the exploration of potentially better configurations.

    \item AIF should more effectively utilize the accumulated experience to minimize negative impact caused by discrepancies in distributions of new and past observations.
\end{enumerate}

It is important to note that our study, particularly experiments for efficiency comparison, are limited to concrete algorithm implementations used. Other libraries or implementation details may impact CPU and memory differently.

\section{Conclusion}\label{sec:conclusions}
This paper benchmarks the AIF method for SLO compliance in DCCS and compares its performance with various RL algorithms, including DQN, A2C, and PPO. 
We focus on adapting to dynamic resource scaling and fluctuating workloads, which often introduce performance and efficiency challenges. 
To evaluate these approaches, we simulate a realistic video conferencing application on an edge device and monitor key metrics such as latency and bandwidth, to ensure service quality by adjusting stream parameters. The experiments incorporate both instantaneous and gradual data shifts, such as network limitations and device overheating, as well as SLO changes to comprehensively assess the adaptability of each algorithm. 
Our results indicate that PPO and A2C achieve high and stable performance across various scenarios, whereas DQN suffers from instability and sample inefficiency.
Meanwhile, AIF demonstrates limited resource consumption and the fastest convergence in a stable scenario, making it a promising approach for DCCS. In the future, we will focus on enhancing AIF for DCCS by planning a series of actions, balancing generative model relevance with computational efficiency, and exploiting accumulated experience to mitigate distribution discrepancies in observations.

\section*{Acknowledgment}
This work is partially funded by the \textit{Svenska Institutet} under the 'SI Baltic Sea Neighbourhood Programme 2024' (Project No. 31005669) and the Swedish Research Council (Project No. 2024-04058).

\section*{Data Availability}
The implementation of our server, agents, and pre-collected metrics are publicly available in the following GitHub repository\footnote{\url{https://github.com/AlfredsLapkovskis/VideoStreamEnv.git}}. The code of our client application is also available at the other repository\footnote{\url{https://github.com/AlfredsLapkovskis/SmartVideoStream.git}}.

\bibliographystyle{IEEEtran}
\bibliography{bibliography.bib}

\balance
\end{document}